\documentclass[11pt]{article}
\usepackage[margin=1in]{geometry}
\usepackage[T1]{fontenc}
\input{glyphtounicode}\pdfgentounicode=1
\usepackage{lmodern,amsmath,amssymb,amsthm,mathtools}
\usepackage[hidelinks]{hyperref}
\usepackage{booktabs,array,tabularx,caption,float}
\usepackage{microtype,titlesec}
\titlespacing*{\section}{0pt}{2.1ex plus .5ex minus .3ex}{1ex plus .2ex}
\titlespacing*{\subsection}{0pt}{1.6ex plus .4ex minus .2ex}{.8ex plus .2ex}
\titlespacing*{\paragraph}{0pt}{1.5ex plus .4ex minus .2ex}{.75em}

\newcolumntype{Y}{>{\raggedright\arraybackslash}X}
\newtheoremstyle{compactplain}{5pt plus 1pt}{5pt plus 1pt}{\itshape}{} {\bfseries}{.}{.5em}{}
\theoremstyle{compactplain}
\hypersetup{pdftitle={Deterministic algorithms for range LIS and its relatives},pdfauthor={Filippo di Gravina and Francesco Damiano Di Gregorio}}
\theoremstyle{compactplain}\newtheorem{theorem}{Theorem}[section]
\newtheorem{lemma}[theorem]{Lemma}

\theoremstyle{remark}

\newcommand{\LIS}{\operatorname{LIS}}

\title{Deterministic algorithms for range LIS and its relatives}
\author{Filippo di Gravina \qquad Francesco Damiano Di Gregorio\\
{\normalsize\textit{Email:} \href{mailto:f.digravina2@studenti.uniba.it}{\texttt{f.digravina2@studenti.uniba.it}},\quad
\href{mailto:f.digregorio16@studenti.uniba.it}{\texttt{f.digregorio16@studenti.uniba.it}}}}
\date{November 29, 2026}

\begin{document}
\setlength{\abovedisplayskip}{5pt plus 2pt minus 2pt}
\setlength{\belowdisplayskip}{5pt plus 2pt minus 2pt}
\setlength{\abovedisplayshortskip}{3pt plus 2pt}
\setlength{\belowdisplayshortskip}{3pt plus 2pt}
\maketitle
\begin{abstract}
We give deterministic algorithms for range longest increasing subsequence (LIS) on a sequence of length $n$ and a batch of $Q$ queries. Write $N=n+Q$ and let $K$ denote the total output size. Two-dimensional range LIS admits reporting in $O((N+K)\log^3(2+n))$ time by composing earliest-endpoint frontiers using totally monotone matrix search. For colored one-dimensional range LIS, compressed hitting sets and monotone min-plus multiplication give $N^{5-\sqrt{13}+o(1)}$ time for the score and attaining color, and $N^{18/13+o(1)}+\widetilde O(K)$ time for reporting. A persistent range oracle gives exact colored and weighted rectangle queries in $\widetilde O(n\sqrt Q+Q)+O(K)$ time. We also obtain a weighted approximation algorithm, algorithms for ordered fragments and tree paths, and reductions relating colored LIS to sparse triangle and masked max-plus problems.
\end{abstract}

\section{Introduction}
A one-dimensional range-LIS query restricts a sequence to an index interval. A two-dimensional query additionally restricts its values, or equivalently asks for a strict increasing chain in a planar rectangle. The colored problem requires a monochromatic chain; the weighted problem maximizes total positive weight. A reporting query returns an optimal chain. Karthik and Rahul~\cite{kr} give randomized reporting algorithms for uncolored 2D, colored 1D and colored 2D with balanced-batch exponents $3/2$, $3/2$ and $5/3$, respectively, plus linear output cost. We improve all three exponents, deterministically, at the expense of logarithmic factors in the output term for the first two problems. For uncolored queries we use the interval oracles of Gawrychowski, Gorbachev and Kociumaka~\cite{ggk}, based on semi-local comparison~\cite{tiskin,tiskin-semilocal}.

\paragraph{Notation.} Let $n,Q\ge1$, $N=n+Q$, $K$ be the total reported cardinality and $\lambda=\log_2(2+n)$. Empty instances are immediate. The notation $\widetilde O$ suppresses polylogarithmic factors. All bounds are total batch times unless a preprocessing or query bound is specified.

\paragraph{Model.} We use a word RAM with polynomially bounded indices and unit-cost key comparisons. For positive real weights we use the exact-arithmetic model of~\cite{korogi}; the approximation algorithm also uses exact multiplication and division. For sequences, rank keys by $(A[i],-i)$; for planar points, rank first coordinates by increasing $x$ and decreasing $y$, then second coordinates by increasing $y$ and decreasing first rank (break identical-point ties consistently). These orders preserve strict chains, and queries include whole original tie groups. For repeated or reversed fragments, use original distinct-value ranks and local traversal order.

Tables~\ref{tab:comparison} and~\ref{tab:relatedcomparison} summarize the algorithms. The colored 1D result uses the monotone product of~\cite{brm} and the rectangular multiplication bounds of~\cite{uz}; the weighted results use the range oracle of~\cite{korogi}.
\begin{table}[H]
\centering\small\setlength{\tabcolsep}{4pt}\renewcommand{\arraystretch}{1.12}
\begin{tabularx}{\linewidth}{@{}>{\raggedright\arraybackslash}p{.31\linewidth}YY@{}}
\toprule
Problem & Previous bound & This work \\
\midrule
Uncolored 1D & $O(n\log^2(2+n)+Q\log(2+n)+K)$~\cite{korogi} & --- \\
Uncolored 2D & $\widetilde O(N^{3/2})+O(K)$~\cite{kr} & $\widetilde O(N+K)$ \\
Colored 1D (reporting) & $\widetilde O(N^{3/2})+O(K)$~\cite{kr} & $N^{18/13+o(1)}+\widetilde O(K)$ \\
Colored 1D (score, color) & $\widetilde O(N^{3/2})$ (score stage) & $N^{5-\sqrt{13}+o(1)}$ \\
\quad Sparse-core alternative & same & $N^{19/4-\sqrt{11}+o(1)}$ \\
Colored 2D & $\widetilde O(N^{5/3})+O(K)$~\cite{kr} & $\widetilde O(N^{3/2})+O(K)$ \\
\bottomrule
\end{tabularx}
\caption{Reporting and score bounds for $Q=\Theta(n)$. The three reporting bounds from~\cite{kr} are randomized; the new bounds are deterministic. The score-stage comparison refers to computation before reporting.}
\label{tab:comparison}
\end{table}

\begin{table}[H]
\centering\small\setlength{\tabcolsep}{4pt}\renewcommand{\arraystretch}{1.12}
\begin{tabularx}{\linewidth}{@{}>{\raggedright\arraybackslash}p{.31\linewidth}YY@{}}
\toprule
Problem & Previous bound or baseline & This work \\
\midrule
Weighted uncolored 1D & $\widetilde O(N)+O(K)$~\cite{korogi} & --- \\
Weighted 2D; weighted colored 1D/2D & $\widetilde O(nQ)$ (elementary) & $\widetilde O(n\sqrt Q+Q)+O(K)$ \\
Weighted 2D, $(1+\varepsilon)$-approx. & --- & $\widetilde O(n+Q/\varepsilon^2)+O(K)$ \\
Ordered tree paths; bounded fragments and edits & $\widetilde O(nQ)$ (elementary) & $\widetilde O(N+K)$ \\
Colored 2D, height $o(\log N)$ & $\widetilde O(N^{5/3})+O(K)$ (unrestricted~\cite{kr}) & $N^{18/13+o(1)}+\widetilde O(K)$ \\
\bottomrule
\end{tabularx}
\caption{Further algorithms. Elementary baselines solve each query separately; the bounded-height comparison is for unrestricted inputs.}
\label{tab:relatedcomparison}
\end{table}

\section{Uncolored 2D range LIS by frontier composition}\label{sec:frontier}
\begin{theorem}\label{thm:main}
A sequence of length $n$ can be preprocessed deterministically in $O(n\lambda^3)$ time and space, after which an arbitrary 2D range-LIS query of optimum length $\ell$ is answered in $O((1+\ell)\lambda^3)$ time, including reporting. The total batch time is $O((N+K)\lambda^3)$.
\end{theorem}

\paragraph{Interval oracles.} For a permutation of length $s$, Gawrychowski, Gorbachev and Kociumaka (GGK)~\cite[Theorem 3 and Lemma 38]{ggk} give $O(s\log^2(2+s))$ preprocessing, $O(\log(2+s))$ length queries and $O(\ell\log(2+s))$ reporting for a nonempty interval. Both structures are deterministic and use space at most their preprocessing time.
\begin{lemma}[Interval anti-Monge inequality]\label{lem:monge}
Let $H(a,b)=\LIS(X[a+1..b])$. For $a\le a'\le b\le b'$,
\begin{equation}\label{eq:monge}
 H(a,b)+H(a',b')\ge H(a,b')+H(a',b).
\end{equation}
\end{lemma}
\begin{proof}
In the planar alignment grid with vertices $(x,y)$, forward horizontal and vertical edges have weight zero and a diagonal edge from $(x-1,X[x]-1)$ to $(x,X[x])$ has weight one. The maximum weight from $(a,0)$ to $(b,s)$ is $H(a,b)$. Optimal paths from $(a,0)$ to $(b',s)$ and from $(a',0)$ to $(b,s)$ intersect because their boundary endpoints are oppositely ordered. Exchanging suffixes at a common vertex proves the inequality.
\end{proof}

\paragraph{Value decomposition.} Build a balanced tree over distinct compressed ranks. At each node $v$, store its indices $P_v$ in increasing order and build the 1D oracles on their locally renumbered ranks. Each element occurs in $O(\lambda)$ nodes; the total construction and space are $O(n\lambda^3)$. A value interval is the disjoint union of $h=O(\lambda)$ canonical nodes $v_1,\ldots,v_h$ in increasing value order. Define
\[
 L_v(u,x)=\LIS((A[i])_{i\in P_v,\ u<i\le x}).
\]
Two binary searches translate endpoints to the local sequence. A score costs $O(\lambda)$ and a witness of length $d$ costs $O((1+d)\lambda)$. Nondecreasing endpoint translation preserves~\eqref{eq:monge} on valid entries.

\paragraph{Earliest-endpoint frontier.} Fix an index interval $[l,r]$. After processing the first $p$ nodes let $D_p(x)$ be the optimum length up to index $x$, and $s_p=D_p(r)$. Define $F_p[0]=l-1$ and $F_p[t]=\min\{x:D_p(x)\ge t\}$ for $1\le t\le s_p$. The initial frontier is $(l-1)$. As a prefix advances by one position, its LIS increases by at most one; hence $F_p$ is strictly increasing and $D_p(F_p[t])=t$.
\begin{lemma}[Frontier transition]\label{lem:dp}
For $F=F_{p-1}$, $s=s_{p-1}$ and $v=v_p$,
\begin{equation}\label{eq:dp}
 D_p(x)=\max_{0\le j\le s:F[j]\le x}\{j+L_v(F[j],x)\}.
\end{equation}
\end{lemma}
\begin{proof}
Every feasible term concatenates an earlier length-$j$ chain ending at $F[j]$ with a chain in the new block after that endpoint; all new values exceed earlier values. Conversely, split an optimum into its $j$ earlier elements and the rest. The earlier part ends no sooner than $F[j]$, so the rest is eligible in the displayed interval. 
\end{proof}

\begin{lemma}[Totally monotone search]\label{lem:tm}
Complete the implicit matrix in~\eqref{eq:dp} by setting
\begin{equation}\label{eq:mat}
 M(j,x)=\begin{cases}j+L_v(F[j],x)&F[j]\le x,\\-j-1&F[j]>x.\end{cases}
\end{equation}
For $i<j$ and $x<y$, $M(i,x)\le M(j,x)$ implies $M(i,y)\le M(j,y)$. Thus maxima and largest maximizing rows of any $c$ ordered columns are found in $O((s+1+c)\lambda)$ time.
\end{lemma}
\begin{proof}
If row $j$ is invalid at $x$, the antecedent is false: row $i$ is either nonnegative or has a larger negative sentinel. Otherwise both rows are valid at $x,y$ and~\eqref{eq:monge} gives $M(j,y)-M(i,y)\ge M(j,x)-M(i,x)\ge0$. This implication holds in every submatrix, so largest maximizing rows are nondecreasing. SMAWK~\cite{smawk} applied to the transpose with rightmost tie-breaking uses $O(s+1+c)$ evaluations. Row zero is always valid, and each entry costs $O(\lambda)$.
\end{proof}

\begin{proof}[Proof of Theorem~\ref{thm:main}]
For each block $p$, first evaluate $D_p(r)=k_p$ by scanning the old frontier. Invert $D_p$ at thresholds $1,\ldots,k_p$ by simultaneous binary searches, with false endpoint $l-1$ and true endpoint $r$. In each round, sort and deduplicate the midpoints and evaluate their maxima by Lemma~\ref{lem:tm}. Each of $O(\lambda)$ rounds costs $O((1+k_p)\lambda)$, including sorting. The resulting endpoints are $F_p[t]$; one final SMAWK pass records a maximizing row $J_p[t]$ for each, and we set $J_p[0]=0$. Since $D_p$ has unit increments,
\[
 J_p[t]+L_{v_p}(F_{p-1}[J_p[t]],F_p[t])=t.
\]
For traceback, start at $t_h=\ell$ and set $t_{p-1}=J_p[t_p]$ in reverse block order. Report the local interval $(F_{p-1}[t_{p-1}],F_p[t_p]]$ when $t_p>t_{p-1}$. These pieces have increasing values and disjoint, ordered index intervals, and their lengths telescope to $\ell$. Reporting costs $O((h+\ell)\lambda)$. As $k_p\le\ell$ and $h=O(\lambda)$, the query time is $O((1+\ell)\lambda^3)$ with $O((1+\ell)\lambda)$ working words. Summation gives the batch bound.
\end{proof}

\section{Ordered fragment composition and related queries}
\begin{theorem}\label{thm:fragments}
A fixed sequence can be preprocessed in deterministic $\widetilde O(n)$ time and space so that an ordered concatenation of $b$ specified contiguous fragments, each optionally reversed and value-restricted, admits exact LIS reporting in $\widetilde O(1+b(1+k))$ time, where $k$ is the optimum cardinality. Repeated fragments are permitted.
\end{theorem}
\begin{proof}
Build a segment tree over original positions. For each canonical block and each traversal direction, sort the points by (original value, negative traversal position), using traversal position as the LIS key in the resulting sequence. An interval of original values corresponds to an interval of this sorted list; its LIS witnesses are exactly the chains in the block under that traversal. Equal values cannot both be selected. Each occurrence appears in $O(\log n)$ blocks, so GGK's structures cost $\widetilde O(n)$ overall.

A query consists of $O(b\log n)$ directed canonical blocks in concatenation order. Let $F[j]$ be the smallest original distinct-value rank of a prefix chain of length $j$, with $F[0]=0$. Each finite positive tail is strictly larger than its predecessor. Let $L_B(u,x)$ be the block's LIS restricted to ranks $(u,x]$ and its prescribed value filter. This is a one-dimensional oracle call; clamping endpoints to the filter is order preserving and retains the valid anti-Monge inequality. The next frontier is the inverse of $\max_{j:F[j]\le x}(j+L_B(F[j],x))$, by the same concatenation argument as Lemma~\ref{lem:dp}. This maximum increases by at most one per original value rank, even with repeated occurrences: a strict chain contains at most one occurrence of the newly allowed value. The completion in Lemma~\ref{lem:tm} and the simultaneous inverse searches apply verbatim, with values instead of indices. Every intermediate length is at most $k$. The time per block is $O((1+k)\lambda^2)$, and traceback reports exactly $k$ occurrence identifiers (including the fragment identifier if repeated). Summing over blocks proves the result.
\end{proof}
\paragraph{Tree paths.} Heavy-light decomposition partitions each directed tree path into $O(\log n)$ oriented fragments, counting the lowest common ancestor once. Preprocess the disjoint heavy paths in both orientations. Theorem~\ref{thm:fragments} gives $\widetilde O(N+K)$ total reporting, also with value filters.
\paragraph{Static sensitivity.} Independent deletion, insertion, reversal and rearrangement queries reduce to fragment queries, with inserted elements represented as singletons and their values included in offline rank compression. With $b_q$ fragments and inserted elements the batch costs $\widetilde O(n+\sum_q(1+b_q(1+k_q)))$. These are independent edits, not a dynamic update sequence.
\paragraph{Sparse substring--substring LCS.} For strings $A,B$, enumerate the $M$ equal-symbol pairs $(i,j)$ from occurrence lists. A common subsequence of substrings corresponds exactly to a strict planar chain of pairs in their rectangle. Sort by first coordinate, breaking ties in decreasing second-coordinate order, and use the 2D theorem with whole tie groups at query boundaries. This gives $\widetilde O(|A|+|B|+M+Q+K)$ reporting, where $M$ may be quadratic in the string lengths. Rectangle-restricted independent sets in permutation graphs are increasing chains in the two permutation orders; cliques are decreasing chains, obtained by reflection. Both inherit the same bound in the permutation representation.

\section{Colored one-dimensional range LIS}\label{sec:colored1}
\subsection{Algebraic primitive and threshold decisions}
By Uzna\'nski~\cite[Theorem 1.1]{uz}, multiplication of an $N^a\times N^r$ matrix by an $N^r\times N^a$ matrix takes $N^{\mu(a,r)+o(1)}$ arithmetic operations, where
\[
 \mu(a,r)=\begin{cases}2a&r\le a/2,\\ a+r+a^2/(4r)&r\ge a/2.\end{cases}
\]
For nonnegative integer operands with each column of the first and each row of the second nondecreasing, range $[0,R]$, inner dimension $r_0$, and $q$ requested entries, the min-plus product of~\cite[Theorem 2.1]{brm} takes
\[
 \widehat O\left(M_0+q+(r_0 R^2 M_0^2)^{1/3}
   +\sqrt{r_0 R\min(R,\sqrt q)M_0}\right),
\]
where $M_0$ includes ordinary multiplication, input and dense output, and $\widehat O$ suppresses subpolynomial factors in $N$. The prime-field implementation of~\cite[Lemma 4.1]{brm} extends from aspect ratios in $[0,1]$ to $[0,3]$: for fixed exponent slack, take a finite mesh in $[0,3]$ and a finite exact bilinear scheme at each mesh point using~\cite[Theorem 1.1]{uz}. The finitely many coefficient identities admit a common number field by the Nullstellensatz. Clear denominators of the schemes and of a basis multiplication table with first basis element $1$. Reduction modulo any sufficiently large prime preserves the identities in a constant-dimensional algebra over the prime field. Embedding the inputs as multiples of $1$ and extracting the first output coordinate gives the required product, with constant overhead for fixed slack. Padding to the next mesh point and continuity give uniform bounds. All primes used by the verifier have polynomial magnitude. Thus, for value range $N^\rho$ and $q\le N^b$, the cost exponent is
\begin{equation}\label{eq:product}
 T(a,r,\rho,b)=\max\left\{\mu(a,r),b,\frac{r+2\rho+2\mu(a,r)}3,\frac{r+\rho+\min(\rho,b/2)+\mu(a,r)}2\right\}.
\end{equation}

Build an interval oracle for each color, with endpoints translated by rank searches. For a fixed threshold $t$, binary search for the first end attaining $t$ at every local start. A query has score at least $t$ iff it contains one of these at most $n$ intervals. Offline dominance tests all queries in $\widetilde O(N)$ time. Testing powers of two locates the scale $v\le\mathrm{OPT}<2v$ of each nonempty query in the same bound.

\subsection{Deterministic compressed hitting sets}
We first give a hitting-set algorithm for compressed families of witnesses.

\begin{lemma}[Compressed hitting sets]
A family of sets represented by target roots of a binary concatenation DAG of size $D$, each root expanding to a list of at least $L$ elements of an $n$-element universe, without repetitions, admits a deterministic hitting set of size $O((n/L)\log(D+1))$ in $\widetilde O(D+n)$ time.
\end{lemma}
\begin{proof}
Remove unreachable nodes, deduplicate targets and identify leaves with the same element. Convert each concatenation bottom-up to a persistent AVL rope by joining its child ropes (copy the boundary spine of the taller tree, joining recursively and restoring balance by rotations). Each join costs $O(\log n)$ new nodes and time. The resulting DAG has near-linear size and depth $H=O(\log n)$; no target can reach a nonempty descendant by two paths, since that would repeat its elements.

For an unhit target indicator $b(u)$, the exact number of unhit roots reaching $u$ is $C(u)=b(u)+\sum_{p\to u}C(p)$. At leaves this is set incidence. Initialize messages $m(u)=C(u)$ by topological evaluation. Maintain $S(u)=b(u)+\sum_{p\to u}m(p)$, and after decreases propagate a new message $m(u)=S(u)$ only if $m(u)>(1+1/(H+1))S(u)$. Process all triggered decreases. The invariant $S(u)\le m(u)\le(1+1/(H+1))S(u)$ implies by induction on depth that $C(u)\le m(u)\le eC(u)$. Messages start at most at the number of roots, decrease geometrically until zero, and have $O((H+1)\log(D+n+1))$ changes. Each node has at most two children, so propagation and maintenance of a leaf max-heap cost near-linear time; comparisons use exact integers.

If $z$ targets remain, their incidences sum to at least $zL$, so the maximum-message leaf hits at least $zL/(en)$ targets. Traverse reverse edges from this leaf to delete those roots, marking visited nodes permanently (all ancestors were visited when a node was first marked). Reverse edges are scanned only once over all deletions. Each choice leaves at most a $1-L/(en)$ fraction of the targets.
\end{proof}
\begin{lemma}[Pivots for long interval witnesses]\label{lem:pivots}
For any integer $v\ge1$, one can construct in $\widetilde O(n)$ time a set of $\widetilde O(n/v)$ positions meeting an optimal monochromatic subsequence of each color interval of LIS length at least $v$.
\end{lemma}
\begin{proof}
Apply Korogi's algorithm~\cite[Algorithm 3 and Lemmas 4.2, 5.1]{korogi} with unit weights separately to the occurrence list of each color. Its saved prefix roots and all allocated binary witness nodes occupy $O(n\log^2(2+n))$ space in total. For each prefix root, a lookup at the left endpoint reads the suffix handles on a root-to-leaf path and concatenates them in reverse order. Every handle at a node reachable from a saved root is part of the witness for some such lookup, so its expansion has no repeated positions. Traverse the union of the reachable function trees and retain the nonempty suffix handles of length at least $v/(\lceil\log_2 n\rceil+1)$, using their stored lengths. Any interval of score at least $v$ has at least one such handle in its optimal witness. Apply the compressed hitting-set lemma to these handles; both the DAG construction and the hitting-set computation take $\widetilde O(n)$ time. The resulting set has size $\widetilde O(n/v)$.
\end{proof}

\subsection{Anchored threshold algorithm}
\begin{theorem}\label{thm:colored1}
Colored 1D range LIS admits deterministic exact score and attaining-color time $N^{5-\sqrt{13}+o(1)}$ and reporting time $N^{18/13+o(1)}+\widetilde O(K)$.
\end{theorem}
\begin{proof}
For each dyadic scale $v=N^{\alpha+o(1)}$, thresholds lie in $[v,2v)$. Choose $d=4-\sqrt{13}$ and $h=1-d$. If $v\le N^d$, enumerate every integer threshold of that scale by the fixed-threshold algorithm above. This costs $\widetilde O(nv+Qv)$, within $N^{1+d+o(1)}$ even if each threshold uses a separate dominance pass; alternatively build all interval records once and test the requested threshold per query. If $v\ge N^h$, only colors of size at least $v$ qualify, and direct evaluation costs $\widetilde O(Qn/v+Q)$.

For intermediate scales, scan colors of size at least $N^h$ per query; there are at most $N^d$. Snap each interval inward to a grid with spacing $D=\lceil N^d\rceil$. Evaluate every color appearing in the discarded boundary positions on the original query. If no interior interval remains, scan the whole query. Every other color has exactly the same occurrences in the snapped interval as in the original interval. Thus these checks and the snapped answers give exact decisions. The overhead is $\widetilde O(N^{1+d})$.

For the remaining colors use Lemma~\ref{lem:pivots} at scale $v$. There are $\widetilde O(n/v)$ pivots, and each interval with score at least $v$ has an optimal witness through a pivot. For a pivot $p$ of color $c$, let $L_p(l)$ be the best length ending at $p$ and using positions at least $l$, and $R_p(r)$ the best length starting at $p$ and using positions below $r$. When $p\in[l,r)$, the best chain through it has length $L_p(l)+R_p(r)-1$. Compute both functions in $\widetilde O(s_c)$ time per pivot by forward/backward Fenwick maximum sweeps on values and prefix/suffix maxima on endpoints. Truncate each side at $2v$; for unavailable sides use $-4v$. Positive thresholds below $2v$ are preserved and invalid pivots cannot qualify. $L_p$ is nonincreasing in $l$, $R_p$ is nondecreasing in $r$. Order the right boundaries decreasingly and complement both operands, giving the required monotonicities and nonnegative range $O(v)$. The exact requested min-plus entries decide the threshold after undoing the complement and subtracting one. If there are no pivots the interior decision is false.

All explicit construction costs $\widetilde O(N^{1-\alpha+h})$; both the grid and the remaining per-color size are bounded by $N^{h+o(1)}$. The product dimensions have outer exponent $h$, inner exponent $1-\alpha$, and value-range exponent $\alpha$. Equation~\eqref{eq:product} yields
\[
 M=h+1-\alpha+\frac{h^2}{4(1-\alpha)},\quad
 T_3=\frac{1+\alpha+2M}3,\quad
 T_2=\frac{1+\min(\alpha,1/2)+M}2.
\]
The smooth branch holds because $d>h/2$. These are exactly the three exponents balanced in~\cite[Section 5.2]{brm}: each is at most $1+d=5-\sqrt{13}$ on $[d,h]$. Construction takes $\widetilde O(N^{2h})$, also within this bound. Parallel binary search within the known dyadic scale gives exact lengths with logarithmic overhead. For color recovery split the palette into two balanced halves, evaluate one half against the known optimum, and route each positive query to a half attaining it. At each depth the total explicit instance size is $O(N)$ after local position compression and query translation, so the sum of superlinear power costs is at most $N^{1+d+o(1)}$. The leaf is a winning color.

For reporting set $d'=5/13$ and first determine the optimum scale $v$ of every query. For $v\le N^{d'}$ use small-scale enumeration. For $v>N^{d'}$ use snapping spacing $\lceil v\rceil$ and scan colors of size at least $N/v$. These two checks cost $\widetilde O(Q_v v)$ for the $Q_v$ queries at scale $v$, chargeable to their actual output length. The remaining color sizes, grid dimensions and pivot count are $\widetilde O(N/v)$. Ordinary square multiplication with exponent $9/4$ gives
\[
 M=\tfrac94(1-\alpha),\quad
 T_3=(1+\alpha+2M)/3,\quad
 T_2=(1+\min(\alpha,1/2)+M)/2.
\]
Each decreases for $\alpha\ge5/13$ on the relevant pieces, and all equal $18/13$ at $5/13$. Explicit construction costs at most $N^{16/13+o(1)}$. The small-scale budget is $N^{1+5/13+o(1)}$. Binary search within the dyadic scale recovers the score at logarithmic overhead. During palette descent retain the original scale and test the known score; a child with fewer than $v$ occurrences is immediately negative. The same analysis with the child's local size applies and the boundary charges sum to $\widetilde O(K)$ per logarithmic depth. Reconstruct only the winning color by the 1D reporting oracle. 
\end{proof}

\subsection{Sparse-core alternative}
\begin{theorem}\label{thm:sparsecore}
A second algorithm for colored 1D range LIS, using sparse mixed differences, has deterministic score and color time $N^{19/4-\sqrt{11}+o(1)}$; reporting adds $\widetilde O(K)$.
\end{theorem}
\begin{proof}
For a color with $s$ occurrences let $F(i,j)=\LIS(X[i..j))$ for $i\le j$, and extend by zero for $i>j$. For each $j<s$, $F(i,j+1)-F(i,j)$ is a nondecreasing binary step for $i\le j$ by Lemma~\ref{lem:monge}, and is zero for $i>j$. Thus the full grid has only $O(s)$ nonzero mixed differences, found by binary search for the first one in each column. On any snapped global grid the local occurrence-prefix maps are monotone, so every global mixed difference is a telescoping sum over a local rectangle. Each local nonzero marks at most one global cell; mark even if cancellation occurs. Across colors the support is $O(n)$ and computable in $\widetilde O(n)$ time. In a tile with no internal mark of a color, summing mixed differences from the corner with largest left and smallest right endpoint shows that its score is $b_c+f_c(l)+g_c(r)$, where the endpoint functions are nonnegative monotone and their variations are bounded by the numbers of occurrences in the corresponding coordinate spans.

\paragraph{Hierarchical evaluation.} Set $d=15/4-\sqrt{11}$ and $h=1-d$. For colors with at most $N^d$ occurrences enumerate every local subinterval, store its score and first/last positions, and use offline dominance maxima, at cost $\widetilde O(N^{1+d})$. Colors with at least $N^h$ occurrences are scanned for every query at the same cost. For the others snap inward to grid spacing $N^d$ and correct colors appearing in the discarded boundary positions exactly as above. Each clean color on a tile has the separable score derived above. Let $w$ be the next power of two above the larger occurrence count on the two coordinate spans, or one if both counts are zero. Within one $w$-group the two functions have range $[0,w]$. For $b_*=\max_c b_c$, discard colors with $b_c<b_*-2w$ and put $b'_c=b_c-b_*+2w$. With left endpoints increasing and right endpoints decreasing, the operands $3w-b'_c-f_c(l)$ and $w-g_c(r)$ satisfy the two monotonicities and have range $O(w)$. The desired maximum is $b_*+2w$ minus the requested minimum. 

Use a dyadic quadtree of the whole snapped grid, retain only tiles with queried cells, and associate the initial root with all remaining colors. At each tile evaluate active clean colors by separability; pass active dirty colors to every nonempty child. Stop at a tile with at most \(N^d\) distinct queried cells and evaluate all active colors individually. The active incidence count at any level is \(O(n)\): at the root it is at most \(n\); otherwise each dirty (parent,color) pair has at least one mark in its parent, so there are \(O(n)\) such pairs at a fixed depth and at most four propagated copies each. Mark classification and query propagation take \(\widetilde O(N)\) total (distribute marks down the quadtree or sort them by ancestor tile). Direct evaluations at terminal tiles cost \(\widetilde O(nN^d)\). Each query-color pair is evaluated once along its path. Clean colors are grouped by local variation \(w\) and normalized as above. If a clean group has at most \(N^d\) factors, evaluate each factor per query: summing over tiles and groups at each depth costs \(\widetilde O(QN^d)\). The \(w=1\) group, regardless of size, can be processed in near-linear incidence plus query time: each of the two monotone integer operand functions has a constant number of jumps, hence each color contributes only constantly many constant-valued rectangles in that tile, found by binary search on the oracle. Maximum-weight rectangle stabbing is near-linear offline via an x sweep and canonical y intervals with heaps of active rectangle weights (insert each rectangle in \(O(\log N)\) nodes and query the maximum over a root-to-leaf path). 

For remaining groups at a fixed level let the tile physical span be \(N^{\gamma+o(1)}\), its outer matrix exponent \(a=\gamma-d\) (if \(\gamma\le d\) it is terminal), group factor count \(N^{r+o(1)}\), queried cell count \(N^{b+o(1)}\), and local variation \(N^{\rho+o(1)}\). We have \(r,b\ge d\), \(r\le h\), \(r+\rho\le\gamma\) by counting occurrences, and \(\rho\le h\). Group tiles in dyadic buckets for \(r,b,\rho\) (polylogarithmically many). There are at most \(N^{1-s+o(1)}\) tiles in a bucket, where \(s=\max(2\gamma-1,r,b)\): use respectively the number of geometric tiles, total active incidences, and total queried cells. Thus \(d\le s\le1\). Pad \(\gamma\) up to \(G=(1+s)/2\), outer dimension to \(A=G-d\), and use \(\rho\le G-r\), \(b\le s\). We need each product term to be at most \(N^{s+d+o(1)}\). Since \(r\ge d>h/2\), the smooth rectangular branch applies:

\[M=A+r+A^2/(4r),\quad T_3=(r+2\rho+2M)/3,\quad T_2=(r+\rho+\min(\rho,b/2)+M)/2.\]

For \(M\) maximize \(r\) at \(R=\min(h,s)\), since \(\partial M/\partial r>0\). If \(s\le h\), \(M-s=A+A^2/(4s)\) is convex in \(s\), so its maximum is at \(d\) or \(h\). If \(s\ge h\), \(M-s=A+h+A^2/(4h)-s\) is decreasing, because its derivative is \(-1/2+A/(4h)<0\). Substituting \(s=d,h\) gives values less than .404, hence less than \(d\).

For \(T_3\), substituting \(\rho\le G-r\) gives \((2G+2A+r+A^2/(2r))/3\), increasing in \(r\) because \(A\le h<\sqrt 2d\). Maximize at \(R\). For \(s\le h\), the upper bound on \(T_3-s\) is \((2-2d+A^2/(2s))/3\), convex in \(s\). For \(s\ge h\) it is \((2-2d+h-s+A^2/(2h))/3\), decreasing. At \(d,h\) the values are less than .414, hence below \(d\). The decimal checks can be made with rational interval \(.4333752096<d<.4333752097\).

For \(T_2\), use the bound \((G+\min(G-r,s/2)+M)/2\). For \(r\le1/2\) the minimum is \(s/2\) and this expression increases with \(r\); for \(r\ge1/2\) the minimum is \(G-r\) and the expression decreases. Since \(h>1/2\), the maximizing \(r\) is \(\min(s,1/2)\). For \(s\le1/2\) the bound on \(T_2-s\) becomes

\[(1+s/2-d+A^2/(4s))/2,\]

which is increasing on \([d,1/2]\) (write \(A=s/2+(1/2-d)\), then the derivative of \(A^2/s\) is \(1/4-(1/2-d)^2/s^2>0\)). For \(s\ge1/2\) the bound is

\[(1.5-d-s/2+A^2/2)/2,\]

which is decreasing since \(A<1\). The maximum at \(s=1/2\) is exactly \(d\): the equation \(d=(1.25-d+(0.75-d)^2/2)/2\) gives \(d=15/4-\sqrt{11}\). (The other root is inadmissible.) Thus each bucket costs at most \(N^{1-s}N^{s+d}=N^{1+d}\) up to logarithms. The additive query term and all explicit input construction are covered. Palette recursion as in the anchored algorithm recovers the winning color. Therefore the sparse-core algorithm has deterministic bound

\[N^{19/4-\sqrt{11}+o(1)}=N^{1.4333752096446\ldots+o(1)}.\]

Reporting uses the attaining color's interval oracle.
\end{proof}

\paragraph{Colored sparse LCS.} For a colored text and a fixed reference string for each color, replace each text occurrence by its matching reference positions in decreasing order, with the same color. A strictly increasing subsequence chooses at most one position from each occurrence and corresponds to a common subsequence of that color. Query endpoints map by prefix lengths. Thus Theorem~\ref{thm:colored1} and the sparse-core fallback transfer with $N$ equal to the explicit match count plus input text/reference lengths and queries; the reported match sequence gives a common subsequence. 

\section{Exact weighted and colored two-dimensional queries}\label{sec:weighted}

\subsection{A persistent three-sided oracle}
\begin{theorem}\label{thm:three}
For weighted chains on $s$ points, any fixed orientation of three-sided rectangle queries admits deterministic preprocessing time and space $O(s\log^2(s+1))$, score-query time $O(\log^2(s+1))$ and reporting time $O(k+\log^2(s+1))$.
\end{theorem}
\begin{proof}
Apply the strict-rank convention of Section~1; identical points of a color may be merged using their heaviest representative. In Korogi's Algorithm 3, process points in increasing first coordinate. Before position $r$, each node $E$ of the outer segment tree on second-coordinate ranks stores the function
\[
 H_E(l)=\max(\{0\}\cup\{D_p(l):p<r,\ y_p\in E\}),
\]
where $D_p(l)$ is the best weight ending at $p$ with first coordinate at least $l$. By~\cite[Lemma 4.1]{korogi}, each merge of value-separated endpoint sets takes $O(\log s)$ time and allocates $O(\log s)$ persistent function and witness nodes. Updating a leaf recomputes $O(\log s)$ ancestors. The input envelopes and their witnesses remain unchanged. Path-copy the $O(\log s)$ modified outer nodes and retain the root of each prefix. The path copies require $O(s\log(s+1))$ additional nodes.

For $[l,r)\times(-\infty,Y]$, use prefix root $r$, decompose the $y$-prefix into $O(\log s)$ canonical nodes and maximize $H_E(l)$. A chain ending at $y_p\le Y$ lies entirely below $Y$. Evaluate functions in $O(\log s)$ time and expand only the winner using Korogi's Lemma 4.2. Empty ranges return zero. Exchanging coordinates and reversing both orders gives the other orientations. In particular, with a fixed lower first-coordinate bound, preprocess the suffix beyond it and exchange coordinates to handle both second-coordinate bounds and a variable upper first-coordinate bound.
\end{proof}

\subsection{Exact colored and weighted rectangles}
\begin{theorem}\label{thm:weightedrect}
Weighted colored rectangle queries on $n$ points take deterministic $\widetilde O(n\sqrt Q+Q)+O(K)$ total time. The score-only bound omits $K$. Unit weights give colored 2D range LIS in $\widetilde O(N^{3/2})+O(K)$ time for balanced batches, without matrix multiplication.
\end{theorem}
\begin{proof}
Set $L=\min(n,\lceil\sqrt Q\rceil)$ and split colors by size.

\emph{Light colors ($s\le L$).} For each first point $p$, run a weighted increasing-subsequence DP in first-coordinate order with a Fenwick maximum tree on second-coordinate ranks, restricting chains to start at $p$. In $O(s\log(s+1))$ time obtain the best weight and predecessor for every possible last point $q$. A chain lies in the rectangle spanned by its endpoints. Store each comparable pair's score, including singletons; an offline four-dimensional dominance maximum finds the best record contained in a query. This costs $\widetilde O(nL+Q)$ in total, and the chosen trace costs $O(k)$ to report.

\emph{Heavy colors ($s>L$).} Partition the first-coordinate list into at most $L$ blocks of size $B_c=\lceil s/L\rceil$. At each block boundary preprocess the suffix with Theorem~\ref{thm:three}, fixing its lower first-coordinate bound. This costs $\widetilde O(sL)$ per color. A rectangle selects an initial strip of at most $B_c$ points, followed by a core from the next block boundary to the query's upper first-coordinate bound. If there is no core, solve the strip directly. Otherwise filter the strip by the query's second-coordinate bounds and compute the best chain weight $d(p)$ ending at each eligible $p$ in $O(B_c\log(s+1))$ time. Query the core oracle for its unrestricted optimum within the rectangle and, for each $p$, for the optimum restricted to $y>y_p$, adding $d(p)$. Take the maximum. A chain either avoids the strip or has a unique last strip point $p$; the two pieces concatenate precisely under this strict inequality. Empty pieces have score zero.

The score cost per heavy color and query is $\widetilde O(B_c)$. Since $B_c\le2s/L$, the total cost is $\widetilde O(nL+Qn/L+Q)$, plus $O(K)$ for the selected witnesses. If $L=n$ there are no heavy colors. Store only the attaining color and record, rerunning its strip DP if necessary within the same budget; expand only the chosen trace and core witness. Substitution of $L$ gives the stated bound.
\end{proof}

\paragraph{Special cases.} One color gives weighted uncolored 2D; unrestricted values give weighted colored 1D; three-sided rectangles are a special case. Uncolored weighted 1D already has total reporting time $O(n\log^2(2+n)+Q\log(2+n)+K)$ by~\cite{korogi}. Weighted colored LCS on explicit compatible pairs reduces to weighted colored chains (the pair-list size counts in $n$), and weighted monochromatic independent sets and cliques in permutation graphs follow by the geometric correspondence of Section~3.

\paragraph{An unweighted alternative.} The interval oracle alone also gives a deterministic $\widetilde O(N^{5/3}+K)$ algorithm. Set $L=\lceil N^{1/3}\rceil$, enumerate light colors as above, and partition each heavy color of size $s$ into blocks of size $B=\lceil s/L\rceil$. Build interval oracles for all $O(L^2)$ unions of consecutive complete blocks, ordered by value with position as key. For a query, compute the best chains ending or starting at points of its two boundary strips (and include empty sentinels). For each feasible pair of strip endpoints, the best chain is the sum of these two scores and the middle oracle's score between their values. In increasing value order this matrix is anti-Monge on its feasible staircase, with row and column offsets. Rightmost row maxima are monotone: a reversal would give two columns feasible in both rows and the anti-Monge inequality would force a tie at the larger column. Divide and conquer finds the maxima in $O(B\log(B+1))$ evaluations. Queries without a complete block are solved directly. Total time is $\widetilde O(nL^2+Qn/L+nL+Q+K)$; reconstruct only the winning chains.

\section{Weighted uncolored approximation}\label{sec:approx}
\begin{theorem}
Uncolored weighted 2D range LIS admits $\widetilde O(n)$ preprocessing and space and $\widetilde O(\varepsilon^{-2})$ query time for a feasible $(1+\varepsilon)$-approximation to the optimum score, for $0<\varepsilon\le1$. Reporting adds $O(k)$, where $k$ is the returned chain length; the batch time is $\widetilde O(n+Q/\varepsilon^2)+O(K)$.
\end{theorem}
\begin{proof}
Build the position-block decomposition of the fragment theorem with Korogi's weighted oracle in place of GGK's. A query uses $B=O(\log n)$ canonical blocks in increasing x order, each filtered by its y-range. Let $F_j(u,y)$ be the optimum weight in block $j$ with second-coordinate ranks in $(u,y]$ intersected with this filter. These scores cost polylogarithmic time; construction is $O(n\log^3(2+n))$. Return the empty chain if the rectangle is empty. Otherwise an orthogonal maximum query gives its largest point weight $W$, with $W\le\mathrm{OPT}\le nW$. If $\varepsilon<1/n$, an exact weighted LIS sweep on the eligible points is within the stated query budget; hence assume $\varepsilon\ge1/n$ below.

Put $\delta=\varepsilon/(4B)$, $\eta=\varepsilon W/(4B)$ and use the grid $G=\{0,\eta,\eta(1+\delta),\ldots\}$ up to $nW$. It has $g=O(1+(B/\varepsilon)\log(nB/\varepsilon))$ elements. Rounding $z\in[0,nW]$ down to this grid loses at most a factor $1+\delta$ and additive $\eta$: the rounded value is at least $z/(1+\delta)-\eta$. The state $T_{j,a}$ for $a\in G$ is the earliest ending rank at which the rounded DP certifies a feasible chain of weight at least $a$ after block $j$; impossible states have value infinity. Initially only $T_{0,0}=0$ is finite, and $T_{j,0}=0$ always. Set
\[
 H_j(y)=\max_{a\in G:T_{j-1,a}\le y}\{a+F_j(T_{j-1,a},y)\}.
\]
All terms are feasible lower bounds, since the earlier witness ends before the new block's selected values. For each positive grid state binary search the smallest $y$ with $H_j(y)\ge a$, scanning at most $g$ earlier states per evaluation. Monotonicity suffices; no weighted Monge property is required. The total is $O(Bg^2\operatorname{polylog} n)$.

Let $L_j(y)$ be the exact optimum and $R_j(y)=\max\{a:T_{j,a}\le y\}$. By induction,
\[
 L_j(y)\ge R_j(y)\ge L_j(y)/(1+\delta)^j-j\eta.
\]
Feasibility gives the first inequality. Split an optimal chain into earlier weight $u$ ending at rank $z$ (zero for the empty chain) and local weight $v$. The state $R_{j-1}(z)$ ends no later than $z$, so the local piece is eligible. Hence $H_j(y)\ge (u+v)/(1+\delta)^{j-1}-(j-1)\eta$. Rounding down proves the induction. At the final rank, $(1+\delta)^{-B}\ge1-B\delta$ and $B\eta=\varepsilon W/4\le\varepsilon\mathrm{OPT}/4$. Therefore the certified score is at least $(1-\varepsilon/2)\mathrm{OPT}\ge\mathrm{OPT}/(1+\varepsilon)$.

For traceback save the maximizing state and block interval at each finite endpoint. Query the reporting oracle only on the chosen intervals; their witnesses concatenate, with actual weight at least the certified score. Per-block overhead is covered by the query bound. This also applies to a fixed number of directed fragments with per-fragment filters, with $B$ the total canonical block count.
\end{proof}

\section{Fine-grained reductions for bounded-height and weighted LIS}
\subsection{Height two and sparse triangle}
All-Edges Sparse Triangle (AE-ST) asks whether each requested pair of vertices in two parts of a tripartite graph has a common neighbor in the third part. The counting version asks for the number of such neighbors. Equivalently, the decision problem is offline set intersection on a prescribed list of pairs. General graph instances reduce to the tripartite form by making three copies of the vertex set and retaining all corresponding ordered edges.

\begin{theorem}\label{thm:height2}
Colored 2D range LIS threshold two is equivalent to AE-ST under near-linear deterministic reductions, even if the global uncolored height is at most two. Counting monochromatic comparable pairs is similarly equivalent to per-edge triangle counting. The decision, count and one-pair witness problems admit $N^{18/13+o(1)}$ time. A feasible approximation of factor strictly less than two also solves the decision problem.
\end{theorem}
\begin{proof}
For the forward reduction let the two query parts be $I=[a]$ and $J=[b]$, and let $k$ range over the third part. Create one left point of color $k$ for each incidence $(i,k)$ and one right point for each incidence $(k,j)$. Choose $S$ larger than the total number of incidences plus one, and use distinct offsets $e\in[1,S-1]$ within each box:
\[
\begin{aligned}
 \mathrm{left}(i,k)&=(Si+e,S(a-i)+S-e),\\
 \mathrm{right}(k,j)&=(S(a+j)+e,S(a+b-j+1)+S-e).
\end{aligned}
\]
All left points form an antichain, all right points form an antichain, and each left point is strictly below and to the left of every right point. For query $(i,j)$, use the rectangle
\[
 [Si,S(a+j)+S-1]\times[S(a-i),S(a+b-j+1)+S-1].
\]
Its lower bounds isolate precisely left box $i$ and its upper bounds isolate precisely right box $j$. A monochromatic pair exists iff the query has a common neighbor; the number of such pairs equals the number of common neighbors. Coordinates have polynomial magnitude and the reduction is linear-size. With optimum at most two, a feasible approximation factor below two distinguishes optimum two from optimum at most one, even for reporting.

For the reverse reduction, build a balanced binary tree over distinct ranks for each coordinate. A point $p$ receives left labels $(c,u,v)$ for all ancestor pairs $u,v$ whose left children contain its first and second rank, respectively, and receives right labels from the right children analogously. Here $c$ is its color. A strict comparable pair $p<q$ of one color shares exactly one left/right label: the two LCAs of their ranks. Every shared label certifies both strict inequalities. This is a biclique partition with $O(n\log^2 N)$ incidences.

For a rectangle with lower corner $l$ and upper corner $r$, select left endpoints $p\ge l$ and right endpoints $q\le r$, with coordinatewise inequalities. A shared label ensures $p<q$ and therefore both endpoints are in the rectangle. Orthogonal range trees decompose the two orthants into disjoint canonical sets; each has a label list of its points. Across the two structures the total list size is $\widetilde O(n)$ and each query induces polylogarithmically many canonical-set pairs. Store a representative for each label. For counts, store multiplicities $a_{S,h},b_{T,h}$ of each label $h$ in canonical sets $S,T$ and sum
\[
 \sum_h a_{S,h}b_{T,h}
\]
over the canonical pairs. The partitions are disjoint and every comparable pair shares exactly one label, so no pair is double counted. Binary expansion of the polynomially bounded multiplicities reduces the weighted dot products to polylogarithmically many unweighted per-edge triangle-count instances.

For completeness, for total incidence size $M$ and $Q$ intersection requests put $P=M+Q$ and $D=\lceil P^{5/13}\rceil$. Scan a smaller set for each query incident to a set of size at most $D$, using dictionaries at cost $\widetilde O(QD)$. At most $M/D$ sets remain. For labels incident to at most $D$ such sets, enumerate pairs of incidences and update only requested pairs, at cost $\widetilde O(MD)$. The number of remaining labels is at most $M/D$. Compute their contributions by ordinary matrix multiplication of dimensions at most $M/D$, including the dense output (cf.~\cite{ayz}). With $\omega\le9/4$, total time is
\[
 \widetilde O(PD)+(P/D)^{9/4+o(1)}=P^{18/13+o(1)}.
\]
Use exact sufficiently large characteristic and the finite-field convention above. This also computes counts. To recover a common label per positive query, recursively split the label universe into halves and evaluate one half per query, routing to a positive half. At each depth total incidence and request sizes are at most the originals; convexity of the power bound preserves it up to logarithms. A leaf identifies the desired label and hence the two representatives. 
\end{proof}

\subsection{Weighted height two and sparse masked max-plus}
Sparse masked max-plus asks for $\max_k(A(i,k)+B(k,j))$ on requested cells $(i,j)$, with missing entries equal to minus infinity. Input size counts all finite entries and requests.
\begin{theorem}
Weighted colored 1D and 2D range LIS on height-at-most-two inputs are each equivalent to sparse masked max-plus under near-linear deterministic reductions. The equivalence holds for polynomially bounded positive integer LIS weights, with witnesses of at most two elements.
\end{theorem}
\begin{proof}
In the reverse reduction of Theorem~\ref{thm:height2}, replace each label multiplicity by the maximum weight of a point with that label. A requested masked max-plus entry then gives the best feasible pair in the corresponding canonical sets. Maximize over canonical pairs, also considering the best singleton (an orthogonal maximum query) and the empty chain. Store the attaining representatives if requested; a maximizing inner index can be recovered by splitting the label universe, as for intersections. This works for the height-two promise, or for the best chain of length at most two on arbitrary inputs. For 1D queries take the full value range.

Conversely suppose finite matrix entries lie in $[-W,W]$, $W\ge0$. Set $L=4W+1$ and $C=L(a+b+2)+3W+1$. Make left groups in increasing $i\in[1,a]$ order, followed by right groups in increasing $j\in[1,b]$ order. Assign strictly decreasing values within each entire section, with all left values below all right values. For each finite incidence of color $k$ use positive weights
\[
\begin{aligned}
 w_{\mathrm{left}}(i,k)&=C-Li+A(i,k),\\
 w_{\mathrm{right}}(k,j)&=C+Lj+B(k,j).
\end{aligned}
\]
Query $(i,j)$ contains left groups $i,\ldots,a$ and right groups $1,\ldots,j$; empty groups are allowed. Put $T=2C-Li+Lj$. A monochromatic pair from the exact groups has weight $T+A(i,k)+B(k,j)\in[T-2W,T+2W]$. Any pair using a later left group or earlier right group has weight at most $T-L+2W<T-2W$. Every singleton has weight at most $C+Lb+W<T-2W$ by the choice of $C$. Thus the optimum is at least $T-2W$ precisely when the requested entry is finite; subtract $T$ to obtain it. The reduction is linear-size, even with global height at most two. A 2D query with the full value range simulates the 1D query.
\end{proof}
Consequently a general weighted colored algorithm of time $N^{\tau+o(1)}+\widetilde O(K)$ gives dense $d\times d$ max-plus multiplication in $d^{2\tau+o(1)}$ time, since $N=\Theta(d^2)$ and $K\le2Q$. Standard distance-product reductions transfer that exponent to APSP~\cite{zwick-apsp}. 

In contrast, unweighted 1D threshold two admits a direct $\widetilde O(N)$ algorithm. For each point $q$, find the rightmost preceding point $p$ of its color with smaller value via a Fenwick maximum sweep on values. The query contains a comparable pair iff it contains one of the recorded intervals $(p,q)$, decided by offline dominance.

\subsection{Reachability through prescribed antichain layers}
Assign each point its longest-chain-ending rank $\ell$ within its color. A dominance DP computes all ranks in $\widetilde O(n)$ time; process first-coordinate ties in batches and query strict second-coordinate prefixes. Equal ranks form antichains, and ranks strictly increase along chains. Fix one color and a list $V_1,\ldots,V_h$ of distinct layers in increasing rank order. A valid path selects exactly one point per listed layer, increasing strictly in both coordinates. Within each layer order points by nondecreasing first and nonincreasing second coordinate.
\begin{lemma}\label{lem:reach}
In $\widetilde O(\sum_i|V_i|)$ time the first-to-last endpoint relation of valid paths can be represented as a disjoint union of bicliques with $\widetilde O(\sum_i|V_i|)$ incidences. A represented path can be reconstructed in $\widetilde O(h)$ time.
\end{lemma}
\begin{proof}
For any point, its successors in the next layer form an interval: the first-coordinate constraint is a suffix and the second-coordinate constraint a prefix. Both endpoints are nondecreasing across nonempty rows in the source-layer order. Compute the intervals by binary searches. Remove inactive points by forward reachability from the first layer, keeping in each subsequent layer the union of intervals from active points. A difference array computes each union, and restricting to active subsequences preserves intervals and monotone endpoints.

Consider relations $X\to Y$ and $Y\to Z$ with monotone interval rows, assuming every $y$ is reached from some $x$ and every $z$ from some $y$. For a row $[a,b]$ in $Y$, discard rows with no successor and consider the first and last nonempty rows within $[a,b]$. Their extremal $Z$ endpoints delimit the composed row. Every $z$ between them belongs to some $Y$ row by assumption. If that row precedes $[a,b]$, monotonicity of right endpoints puts $z$ in the first nonempty row of $[a,b]$; if it follows, monotonicity of left endpoints puts $z$ in the last; otherwise it is directly reachable. Thus the composed row is an interval. Its endpoints are nondecreasing as the source interval slides. Applying this fact through the forward-active layers proves that reachable final endpoints form an interval for every active point, or are empty if no suffix exists.

Compute extremal reachable final ranks backward rather than materializing repeated compositions. Final points have singleton intervals. For each earlier point, use sorted lists to find the first and last eligible successor with nonempty final interval, which determine its extremal reachable final ranks. This takes $O(\log n)$ per point. For traceback, among eligible successors with nonempty intervals find the first whose right endpoint reaches the desired final rank, and check its left endpoint. Monotonicity and the interval property guarantee success; repeat through the layers.

Decompose each first-layer point's final interval into disjoint canonical nodes of a binary segment tree over the final active layer. Give that source the corresponding node labels and each final point its ancestor labels. Each connected endpoint pair shares exactly one label. Include color and layer-list identifiers. The total incidences are $O((|V_1|+|V_h|)\log n)$; for $h=1$, simply give each point its own identity label. 
\end{proof}

\subsection{Bounded height and global optimum preservation}
For each color $c$ with $n_c$ points and $H_c$ global layers, apply Lemma~\ref{lem:reach} to every nonempty subset of layers in increasing order and tag the labels by list length. A chain of length $t$ follows exactly one list of $t$ distinct layers. A valid path is contained in the query rectangle iff its first endpoint dominates the lower corner and its last endpoint is dominated by the upper corner. Thus a threshold $t$ query tests set intersections on canonical orthant pairs for lists of length exactly $t$. Separate range-tree structures for each $t$ cost $\widetilde O(\sum_c n_c2^{H_c})$ total incidences. Parallel binary search over $t$ determines all exact lengths; a successful label identifies the color, endpoints and list for reconstruction. Writing
\[
 P=Q+\sum_c n_c2^{H_c},
\]
the result is $P^{18/13+o(1)}+\widetilde O(K)$ for $\max_c H_c=O(\log N)$. In particular $\max_c H_c=o(\log N)$ gives $N^{18/13+o(1)}+\widetilde O(K)$. For every fixed height bound at least two this is equivalent to AE-ST up to near-linear reductions (the threshold searches are over a constant number of layers).

For arbitrary $H_c$, a different problem asks for the color of largest global height that retains a full-height chain inside the rectangle. Use only the list of all layers of each color; its representation is $\widetilde O(n)$ across colors. Add a range-tree dimension for the label score $H_c$, so tests can restrict to $H_c\ge t$. Binary search and witness recovery take $N^{18/13+o(1)}+\widetilde O(K)$, returning none if no global optimum survives. 

For one uncolored point set the preservation predicate is simpler. The lower corner selects an interval of first-layer points, and the upper corner selects an interval of active last-layer points in their antichain orders. In the monotone reachability representation, find the first eligible nonempty row whose right endpoint reaches the target interval's left endpoint, then test whether its left endpoint is at most the target interval's right endpoint. Binary searches give $O(\log n)$ query time after $\widetilde O(n)$ preprocessing, and successful reporting takes $\widetilde O(H)$ for global height $H$.

\paragraph{A dense reduction with globally increasing values.} For integer $m\times m$ matrices $A,B$, create one color for each inner index $k$. In left groups $i=1,\ldots,m$ put one element of each color with weight $C+A_{ik}-A_{i+1,k}$, and in right groups $j=1,\ldots,m$ use $C+B_{kj}-B_{k,j-1}$, with zero boundary rows/columns and sufficiently large $C$. Assign values equal to global positions. The interval from left group $i$ to right group $j$ has color-$k$ optimum $C(m-i+1+j)+A_{ik}+B_{kj}$ by telescoping. Subtract the known offset to recover the dense max-plus entry. This proves the score-only transfer for globally increasing inputs of size $\Theta(m^2)$. The witnesses in this construction can have length $\Theta(m)$.

\section{Crossing covers}

Let \(P\) be \(s\) points sorted by position, \(m\) a cut, \(1\le k_1\le k_2\), and \(F(a,r)\) the strict LIS length of the half-open interval \([a,r)\). A crossing cover (cf.~\cite{gj}) is required to provide, for every \(a\le m\le r\) with \(F(a,r)\ge k_2\), a contained interval and an explicit increasing witness of length at least \(k_1\). Covers can include a representative on each side of the cut. A witness interval may be shrunk to the actual first and last witness positions.

\begin{lemma}[Crossing covers]\label{lem:cover} Such a cover can be constructed in \(\widetilde O(s+k_1+k_1^2/(k_2-k_1+1))\) time, using \(O(1+k_1/(k_2-k_1+1))\) witnesses. At most two witnesses are wholly on one side, and the crossing witnesses can be chosen with neither interval containing another. If \(s<k_2\), the empty cover suffices.
\end{lemma}
\begin{proof} Preprocess the sequence with the GGK interval oracle. Let \(a_0\) be the latest start with \(F(a_0,m)\ge k_1\), or \(-1\) if none. Let \(r_0\) be the earliest end with \(F(m,r_0)\ge k_1\), or \(s+1\) if none. Binary searches find both; report one length-\(k_1\) witness per existing interval. Every crossing query with \(a\le a_0\) or \(r\ge r_0\) is covered already. The remaining domain has \(a_0<a\le m\le r<r_0\), and both single-side scores are less than \(k_1\).

Starting at \(a=a_0+1\), find the least \(r\in[m,r_0)\) with \(F(a,r)\ge k_2\). If there is none, terminate: increasing the left boundary cannot improve the score. Find the greatest \(b\ge a\) such that \(F(b,r)\ge k_1\). Since \(F(m,r)<k_1\), we have \(b<m\). Report a length-\(k_1\) witness in \([b,r)\) (truncate an optimum if needed). This covers every eligible crossing query with starting boundary in \([a,b]\): its right boundary is at least \(r\), since \(r\) was minimal at the earlier start \(a\). Continue at \(a'=b+1\). Each step uses binary searches and one output-sensitive witness query. The LIS anti-Monge inequality (Lemma~\ref{lem:monge}) gives

\[F(a,m)+F(a',r)\ge F(a,r)+F(a',m),\]

hence \(F(a,m)-F(a',m)\ge k_2-k_1+1\), as \(F(a',r)<k_1\). The initial left score is below \(k_1\) and nonnegative; the step bound follows. The selected witnesses must cross the cut in the remaining domain. Successive start bounds increase. Their actual ending boundaries increase too, since no witness of length \(k_1\) starting at or after \(a'\) can end at or before the preceding \(r\). Shrinking to actual witness endpoints preserves coverage and noncontainment. \end{proof}

\section{Weighted fragment queries and 3SUM}
For related reductions from 3SUM, see~\cite{patrascu,kpp}.
\begin{lemma}\label{lem:intersection}
Let $X,Y\subseteq[1,U]$. In descending label order form a left fragment with keys $2x$ and weights $x+1$, and a right fragment with keys $2y+1$ and weights $U+1-y$. Their concatenation has weighted LIS at most $U+2$, with equality if and only if $X\cap Y\ne\varnothing$; every witness has at most two elements.
\end{lemma}
\begin{proof}
Both fragments decrease, and a cross pair is feasible exactly if $x\le y$. Its weight is $U+2+x-y$, attaining $U+2$ precisely for equal labels. Singletons weigh at most $U+1$.
\end{proof}
\begin{theorem}[Independent preprocessing barrier]
On a deterministic word RAM with $O(\log s)$-bit words, suppose the two size-$s$ sets in Lemma~\ref{lem:intersection} are independently preprocessed, with input-independent common information and polynomially addressable representations. Some exact query needs $\Omega(s/\log s)$ word accesses, even though its LIS cardinality is at most two.
\end{theorem}
\begin{proof}
Set $U=2s$. Alice and Bob locally construct their respective representations. Each adaptive read can be simulated by its owner sending the word, since the address and control state are known from the transcript and public parameters. $T$ word accesses give a deterministic $O(T\log s)$-bit disjointness protocol. The disjoint input pairs $(S,[2s]\setminus S)$ for all size-$s$ subsets $S$ form a fooling set: distinct such pairs have a cross pair with nonempty intersection. No disjoint transcript rectangle can contain two of them. There are $\binom{2s}{s}$ pairs, so $\Omega(s)$ bits are necessary. The lower bound concerns independently preprocessed inputs.
\end{proof}
\begin{theorem}[Reduction from integer 3SUM]\label{thm:3sum}
Suppose a collection of descending weighted fragments of total length $M$ can be jointly preprocessed in deterministic $M^{p+o(1)}$ time, $p\ge1$, to answer exact weighted LIS of any ordered pair of contiguous subfragments in $\widetilde O(1+s^\kappa)$ time, where $s$ is their combined length and $0\le\kappa\le1$. Then three-role integer 3SUM on polynomially bounded integers admits deterministic time $n^{E+o(1)}$ for
\[
 E=\max\{3p/2,(3+\kappa)/2\}.
\]
\end{theorem}
\begin{proof}
Consider three sets $A,B,C$ of at most $n$ integers, with target $a+b=c$. A standard separating family of role assignments of polylogarithmic size handles distinct input positions. Deduplicate within each role. Put $u=3/2$ and let $r$ be a multiple of three of order $\sqrt{\log n/\log\log n}$. Choose $r$ distinct primes greedily from $[P,2P)$, where $P=\lceil C_0\log n\, n^{u/r}\rceil$ for a sufficiently large constant $C_0$. At each step minimize the total count of triples modulo the product of the previously selected primes and the trial prime. Trial counts are exact cyclic convolutions of residue histograms, summed at residues of $C$. Any nonzero polynomially bounded difference $a+b-c$ is divisible by at most $O(\log n/\log P)$ primes of the pool; the pool has $\Omega(P/\log P)$ primes, of which only $r$ are used. Thus each greedy step reduces the surviving false count by at least a factor $n^{-u/r}$ in the sense that the new count is at most that fraction of the old one. True triples survive every step. The final false count is at most $n^{3-u}$ and the modulus is $n^{u+o(1)}$. There are $n^{o(1)}$ trials, with total time $n^{u+o(1)}$ by exact polynomial convolution and sieving. If total modular count exceeds $n^{3-u}$, return YES. Otherwise denote the count by $Z\le n^{3-u}$.

Divide the primes into three equal groups with products $m_1,m_2,m_3=n^{u/3+o(1)}$, and use the Chinese remainder coordinates in $G=\mathbb Z_{m_1}\times\mathbb Z_{m_2}\times\mathbb Z_{m_3}$. Create three vertex parts $U,V,W$ on the coordinate planes $x_1=0,x_2=0,x_3=0$, respectively. For every $a\in A$, make all edges $UV$ whose difference $V-U$ equals $a$ in $G$, and similarly edges $VW$ for $b$ and $UW$ for $c$. The numbers of such edges per original element are $m_3,m_1,m_2$, respectively. Deduplicate edges, retaining pointers to original residue buckets. The total edge size is $M=O(n(m_1+m_2+m_3))=n^{3/2+o(1)}$. Every modular triple gives the unique triangle
\[
 U=(0,-a_2,-c_3),\quad V=(a_1,0,-b_3),\quad W=(c_1,b_2,0),
\]
and every triangle gives modular triples on expanding its three buckets. Thus expansion costs $O(Z)$ over all triangles.

For each $U$ store its distinct $V$-neighbor labels $X_U$, and for each $W$ its labels $Y_W$, in one common order. Make the two types of fragments using Lemma~\ref{lem:intersection} and preprocess them jointly. The total fragment length is $O(M)$. For every query edge $UW$ we must find all common labels of $X_U,Y_W$. There is an aggregate length bound: for fixed $c$, its $m_2$ edges indexed by $\alpha\in\mathbb Z_{m_2}$ have endpoints $(0,-\alpha,-c_3)$ and $(c_1,c_2-\alpha,0)$. Their list lengths are bounded by $|\{a:a_2=\alpha\}|$ and $|\{b:b_2=c_2-\alpha\}|$, respectively. Summing over $\alpha$ and then over $c$ gives total at most $2n^2$, even before query-edge deduplication.

Use the weighted oracle to test each pair of lists. For a positive test, split by a common median label of their combined ordered lists, sending equal labels to the same child. Both children are contiguous subfragment pairs, each of size at most $\lceil s/2\rceil+1$ because each label occurs at most twice. Stop at constant size and inspect explicitly. Query only nonempty pairs and recurse only on positive ones. Every positive node contains a common label; disjoint nodes at a depth within a query have disjoint labels. The total number of common labels over all query edges is at most $Z$. At each depth there are $O(M+Z)$ tests of total combined length at most $2n^2$. By concavity,
\[
 \sum s^\kappa\le O(M+Z)^{1-\kappa}(2n^2)^\kappa
 =n^{(3+\kappa)/2+o(1)}.
\]
There are logarithmically many depths; medians, splits and original verification add only polylogarithmic overhead or $O(Z)$. Preprocessing costs $M^{p+o(1)}$. Check expanded triples against the original integers, returning YES iff one satisfies equality. 
\end{proof}
In particular, $p=1$ and $\kappa=3/5$ give a transfer exponent of $9/5$. The independent-preprocessing lower bound does not apply to joint preprocessing.

\paragraph{Use of generative AI.} Generative AI tools assisted with literature searches, drafting and editing. The authors take responsibility for all mathematical claims and conclusions.

{\small
\begin{thebibliography}{99}
\setlength{\itemsep}{1pt}\setlength{\parskip}{0pt}
\bibitem{tiskin} Alexander Tiskin. Fast distance multiplication of unit-Monge matrices. In \emph{Proceedings of the 21st Annual ACM-SIAM Symposium on Discrete Algorithms (SODA)}, pp. 1287--1296, 2010.
\bibitem{ggk} Pawe\l{} Gawrychowski, Egor Gorbachev, and Tomasz Kociumaka. Core-Sparse Monge Matrix Multiplication: Improved Algorithm and Applications. In \emph{33rd Annual European Symposium on Algorithms (ESA 2025)}, LIPIcs, 74:1--74:17, 2025. \href{https://doi.org/10.4230/LIPIcs.ESA.2025.74}{doi:10.4230/LIPIcs.ESA.2025.74}; full version \href{https://arxiv.org/abs/2408.04613v2}{arXiv:2408.04613v2} (Theorem 3 and Appendix B).
\bibitem{kr} Karthik C. S. and Saladi Rahul. Range Longest Increasing Subsequence and Its Relatives. In \emph{17th Innovations in Theoretical Computer Science Conference (ITCS 2026)}, LIPIcs, article 87, 2026. \href{https://doi.org/10.4230/LIPIcs.ITCS.2026.87}{doi:10.4230/LIPIcs.ITCS.2026.87}; \href{https://arxiv.org/abs/2404.04795}{arXiv:2404.04795}.
\bibitem{smawk} Alok Aggarwal, Maria M. Klawe, Shlomo Moran, Peter Shor, and Robert Wilber. Geometric Applications of a Matrix-Searching Algorithm. \emph{Algorithmica}, 2:195--208, 1987. \href{https://doi.org/10.1007/BF01840359}{doi:10.1007/BF01840359}.

\bibitem{brm} Filippo di Gravina. Batch range mode in $O(n+Q)^{1.395}$. Preprint, \href{https://arxiv.org/abs/2610.24273v1}{arXiv:2610.24273v1}, 2026.
\bibitem{korogi} Keisuke Korogi. Near-Linear Preprocessing for Range Weighted Longest Increasing Subsequence Queries. Preprint, \href{https://arxiv.org/abs/2610.20983v1}{arXiv:2610.20983v1}, 2026.
\bibitem{uz} Przemys\l{}aw Uzna\'nski. Rectangular matrix multiplication from shared-leg entropy. Preprint, \href{https://arxiv.org/abs/2610.10494v1}{arXiv:2610.10494v1}, 2026.
\bibitem{tiskin-semilocal} Alexander Tiskin. Semi-local string comparison: algorithmic techniques and applications. \emph{Mathematics in Computer Science}, 1(4):571--603, 2008. \href{https://doi.org/10.1007/s11786-007-0033-3}{doi:10.1007/s11786-007-0033-3}.
\bibitem{ayz} Noga Alon, Raphael Yuster, and Uri Zwick. Finding and Counting Given Length Cycles. \emph{Algorithmica}, 17(3):209--223, 1997. \href{https://doi.org/10.1007/BF02523189}{doi:10.1007/BF02523189}.
\bibitem{zwick-apsp} Uri Zwick. All Pairs Shortest Paths using Bridging Sets and Rectangular Matrix Multiplication. \emph{Journal of the ACM}, 49(3):289--317, 2002. \href{https://doi.org/10.1145/567112.567114}{doi:10.1145/567112.567114}.
\bibitem{gj} Pawe\l{} Gawrychowski and Wojciech Janczewski. Fully Dynamic Approximation of LIS in Polylogarithmic Time. In \emph{Proceedings of the 53rd Annual ACM SIGACT Symposium on Theory of Computing (STOC)}, pp. 654--667, 2021. \href{https://arxiv.org/abs/2011.09761}{arXiv:2011.09761}.
\bibitem{patrascu} Mihai P\v{a}tra\c{s}cu. Towards polynomial lower bounds for dynamic problems. In \emph{Proceedings of the 42nd ACM Symposium on Theory of Computing (STOC)}, pp. 603--610, 2010. \href{https://doi.org/10.1145/1806689.1806772}{doi:10.1145/1806689.1806772}.
\bibitem{kpp} Tsvi Kopelowitz, Seth Pettie, and Ely Porat. Higher Lower Bounds from the 3SUM Conjecture. In \emph{Proceedings of the 27th Annual ACM-SIAM Symposium on Discrete Algorithms (SODA)}, pp. 1272--1287, 2016. \href{https://doi.org/10.1137/1.9781611974331.ch89}{doi:10.1137/1.9781611974331.ch89}.
\end{thebibliography}}
\end{document}